\documentclass[reprint,aps,prm,superscriptaddress,showpacs,longbibliography,floatfix]{revtex4-2}

\usepackage{units}
\usepackage{subfigure}
\usepackage{xspace,textcomp}
\usepackage{amsbsy,amssymb,amsmath}
\usepackage{graphicx,color,epsfig,rotate}
\usepackage{upgreek}
\usepackage{tabularx}
\usepackage{setspace}

\usepackage{hyperref}

\begin{document}

\title[Fermi surface and effective masses of IrO\texorpdfstring{$_2$}{2}]{Fermi surface and effective masses of IrO\texorpdfstring{$_2$}{2} probed by de Haas-van Alphen quantum oscillations}

\author{K.~G\"{o}tze}
\affiliation{Deutsches Elektronen-Synchrotron DESY, Notkestrasse 85, 22607 Hamburg, Germany.}
\affiliation{Department of Physics, University of Warwick, Coventry CV4 7AL, UK.}
\author{M.~J.~Pearce}
\altaffiliation{Present address: Department of Physics, Durham University, Durham DH1 3LE, UK.}
\affiliation{Department of Physics, University of Warwick, Coventry CV4 7AL, UK.}

\author{S.~Negi}
\affiliation{Department of Physics, Faculty of Science, National University of Singapore, Science Drive 3, Singapore 117551, Singapore}

\author{J.-R.~Soh}
\affiliation{Quantum Innovation Centre (Q.InC), Agency for Science Technology and Research (A*STAR), 2 Fusionopolis Way, Innovis \#08-03, Singapore, 138634 Singapore}

\author{D.~Prabhakaran}
\affiliation{Clarendon Laboratory, Department of Physics, University of Oxford, Parks Road, Oxford OX1 3PU, UK.}

\author{P.~A.~Goddard}
\email{p.goddard@warwick.ac.uk}
\affiliation{Department of Physics, University of Warwick, Coventry CV4 7AL, UK.}

\begin{abstract}

 Iridium-containing conducting materials are widely investigated for their strong spin-orbit coupling and potential topological properties. Recently the commonly used electrode material iridium dioxide was found to host a large spin-Hall conductivity and was shown to support Dirac nodal lines. Here we present quantum-oscillation experiments on high-quality IrO$_2$ single crystals using the de Haas-van Alphen effect measured using torque magnetometry with a piezo-resistive microcantilever as well as density functional theory-based band-structure calculations. The angle, temperature and field dependencies of the oscillations and the calculated band dispersion provide valuable information on the properties of the charge carriers, including the Fermi-surface geometry and electronic correlations. Comparison of experimental results to calculations allows us to assigns the observed de Haas-van Alphen frequencies to the calculated Fermi surface topology. We find that the effective masses of IrO$_2$ are enhanced compared to the rest electron mass $m_e$, ranging from 1.9 to 3.0~$m_e$, whereas the scattering times indicate excellent sample quality.  We discuss our results in context with recent ARPES and band-structure calculation results that found Dirac nodal lines in IrO$_2$ and compare the effective masses and other electronic properties to those of similar materials like the nodal chain metal ReO$_2$ in which Dirac electrons with very light effective masses have been observed.

\end{abstract}


\maketitle

\section{Introduction}
5$d$ transition-metal oxides are building blocks for a number of compounds whose properties are strongly influenced by spin-orbit coupling (see \cite{witczak-krempa_correlated_2014} and \cite{rau_spin-orbit_2016} for recent reviews).
The dioxide IrO$_2$ in particular has received renewed attention after the discovery of a large spin Hall effect \cite{fujiwara_5d_2013} and the subsequent prediction of Dirac nodal lines (DNL) \cite{sun_dirac_2017}.
ARPES measurements on IrO$_2$ \cite{das_role_2018,xu_strong_2019,nelson_dirac_2019} have confirmed the predictions about the band structure from \cite{sun_dirac_2017} and revealed the importance of spin-orbit coupling and band topology for explaining the spin Hall effect.
IrO$_2$ is commonly used for the production of electrodes, resistors and capacitors in sensors \cite{igarashi_submicron_2000,tankiewicz_new_2001,tilley_light-induced_2010,nakamura_preparation_1994} and is now discussed as a material for applications in spintronic devices \cite{fujiwara_5d_2013}. It is also an important reference material for systems that comprise 5$d$ transition-metal oxide octahedra like IrO$_6$ and that potentially host novel quantum phases \cite{Kawasaki_2016,pesin_mott_2010}.

Quantum oscillation results and band-structure calculations were first reported in the 1970s \cite{graebner_magnetothermal_1976,mattheiss_electronic_1976} but a complete analysis was not possible at the time as the magnetothermal oscillation measurements did not allow for the determination of effective masses and scattering rates.

In this paper we, therefore, present an update on the Fermi surface (FS) properties and electronic correlations of IrO$_2$ including the effective masses and Dingle temperatures. Our results are further compared with new first-principle calculations of the band-structure and are placed in context with previous calculations and with the FS and electronic properties of other topological Dirac materials.

Another aim of this work is to provide a reliable reference for experiments on the band structure of other materials that contain iridium oxide building blocks. Quantum oscillations are yet to be observed in, for example, pyrochlore iridates and their unambiguous identification requires knowledge of the FS topology, corresponding quantum oscillation frequencies and effective masses of closely related materials such as IrO$_2$ that might form in small quantities as a by-product of the growth process.

\section{Methods}

\begin{figure}[t]
	\begin{center}
		{\includegraphics[width=.99\columnwidth]{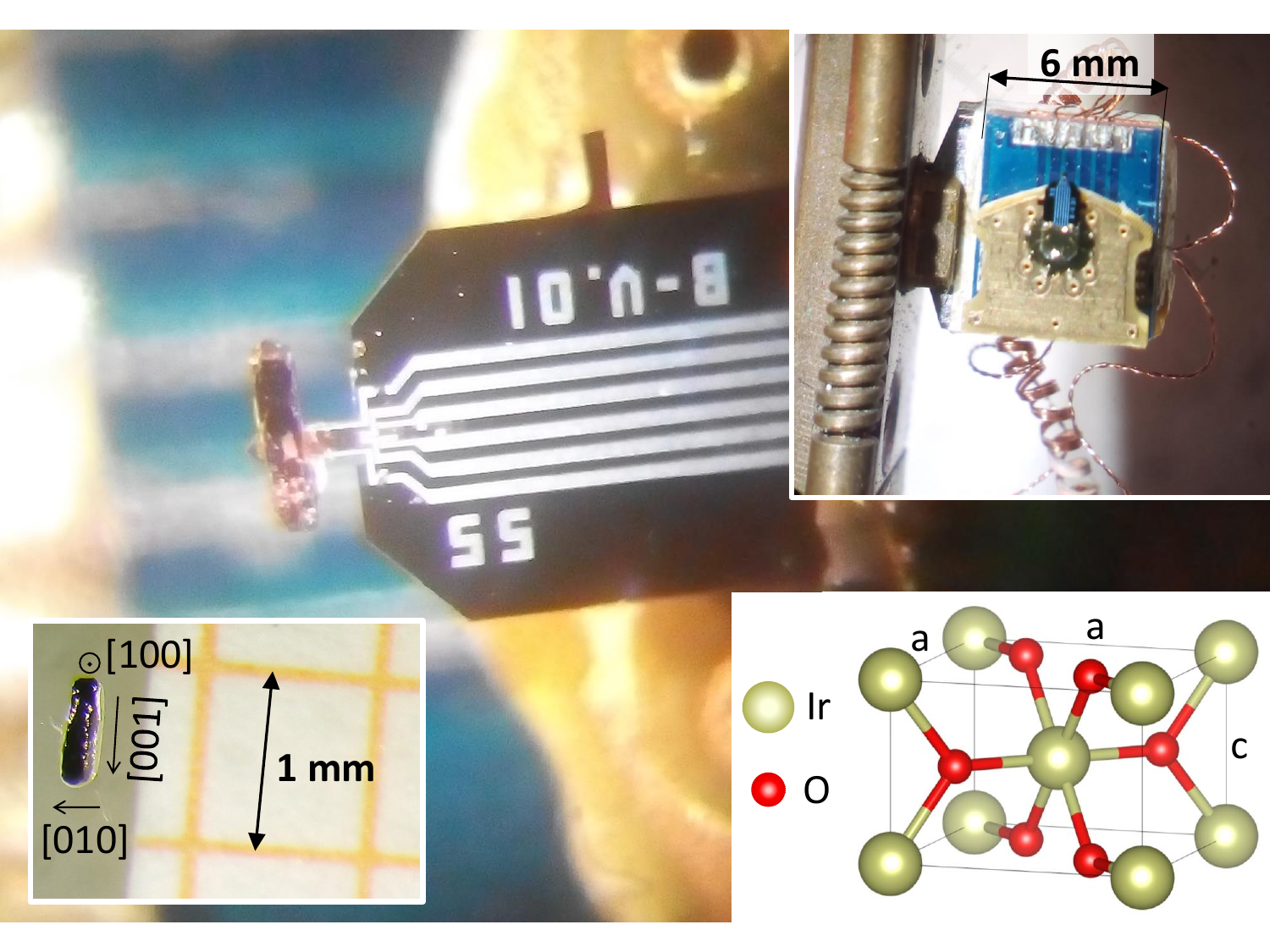}}
	\end{center}
	\vspace{-0mm}
	\caption[IrO2 crystal on microcantilever.]{Main figure: IrO$_2$ crystal on a microcantilever tip. Upper right inset: Microcantilever on PCB counterpart (blue) which, in turn, is glued to the rotator platform. Lower left inset: The same crystal from the main panel next to a piece of scale paper. Lower right inset: Rutile crystal structure of IrO$_2$. This representation of the structure was created using the program \textsc{VESTA} \cite{momma_vesta_2011}.}
	\vspace{-0mm}
	\label{fig:xtal}
\end{figure}

IrO$_2$ crystallizes in a tetragonal rutile structure shown in the lower right inset of Fig.~\ref{fig:xtal} with lattice parameters $a=4.4990$~\AA~and $c=3.1546$~\AA~\cite{rogers_crystal_1969}.
Very similar values are reported in \cite{boman_precision_1970,graebner_magnetothermal_1976,yen_growth_2004}.
IrO$_2$ single crystals were grown by chemical vapor transport as described in \cite{reames_1976}.  High purity iridium metal powder (99.9~\%) was kept in a tube furnace and the oxygen gas flow was set to 20~cc/min. The furnace was slowly heated up to 1050$^{\circ}$C and kept at this temperature for one week.  Small crystals then formed near the cold end (950$^{\circ}$C) of the furnace.

We conducted magnetic torque measurements on IrO$_2$ single crystals in magnetic fields up to 15~T and at constant temperatures between 540~mK and 4.2~K. An IrO$_2$ crystal (lower left inset of Fig.~\ref{fig:xtal}) was attached to the tip of a SCL-Sensor.Tech.~piezo-resistive microcantilever using Apiezon~N vacuum grease as shown in the main panel of Fig.~\ref{fig:xtal}. The crystal was oriented using natural growth faces with the crystallographic orientation indicated in the lower left inset of Fig.~\ref{fig:xtal}. The tip of the cantilever used in this study is 300~$\mu\mathrm{m}$ long, 100~$\mu\mathrm{m}$ wide and about 10~$\mu\mathrm{m}$ thick \cite{SCL}. It is robust enough to allow for repeated mounting of a crystal with the help of a thin, grease covered hair and was found to carry crystals of a few 100~$\mu\mathrm{m}$ edge length without breaking. The cantilever chip is pre-mounted on a PCB (upper inset of Fig.~\ref{fig:xtal}) customized to fit on our 7~mm by 7~mm rotator platform. The rotator mechanism allows for angle-dependent measurements with \textit{in-situ} change of angle.
The piezoresistor that indicates the change of torque is incorporated into a Wheatstone bridge on the cantilever chip. Four-point resistance measurements were conducted with a standard lock-in amplifier, using an excitation current of $\approx$100~$\mu$A.

The Quantum Espresso package \cite{giannozzi2009} was used for first principles calculations with a plane-wave basis set. A kinetic energy cutoff of 50~Ry was used along with fully-relativistic Ultra-Soft pseudopotentials which accounted for the core electrons \cite{rkkjus,dalcorsoO2014} together with the exchange-correlation functional of Perdew, Burke and Ernzerhof \cite{pbe1996}. The Brillouin zone (BZ) was sampled using a $\Gamma$-centered 6$\times$6$\times$6 Monkhorst-Pack grid \cite{mpgrid} for all calculations.

\section{Results}

\begin{figure*}[t]
	\begin{center}
		{\includegraphics[width=.98\textwidth]{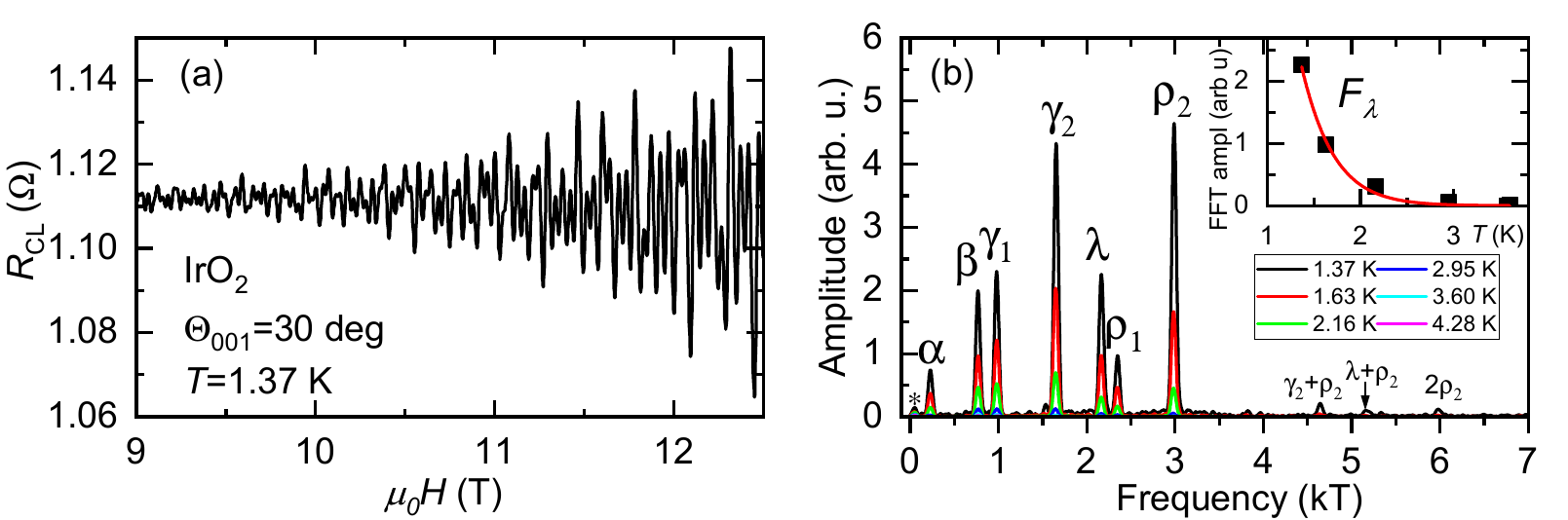}}
	\end{center}
	\vspace{-0mm}
	\caption[IrO$_2$ raw data, FFT at different temperatures and LK fit for 2.16~kT.]{(a) Field dependence of the microcantilever resistance for IrO$_2$ at 1.37~K. (b) Frequency spectrum obtained by fast Fourier transformation (FFT) of the oscillating part of the magnetic torque signal between 9 and 12.5~T at different temperatures between 1.37 and 4.28~K. Fundamental frequencies that were observed before are labeled with the same Greek letters as in \cite{graebner_magnetothermal_1976}. Frequency F*$\approx$50~T has not been reported before. Inset: temperature dependence of FFT amplitudes for $F_\lambda$=2.16~kT and fit line determined by use of the Lifshitz-Kosevich formula yielding $m^*_\mathit{\lambda}=2.3(5)~m_e$, where $m_e$ is the electron rest mass.}
	\vspace{-0mm}
	\label{fig:IrO2_example}
\end{figure*}

We have observed de Haas-van Alphen (dHvA) quantum oscillations in IrO$_2$ which, at 0.54~K, start from about 6~T.
At very high fields and very low temperatures, the oscillations acquire a saw-tooth like aspect indicative of magnetic and/or torque interaction \cite{shoenberg,bergemann_quantum_1999}. (See also Appendix~A.)
In this paper we only consider those fields and temperatures at which purely sinusoidal oscillations are observed and where interaction effects can be excluded.

\begin{figure}[t]
	\begin{center}
		{\includegraphics[width=.98\columnwidth]{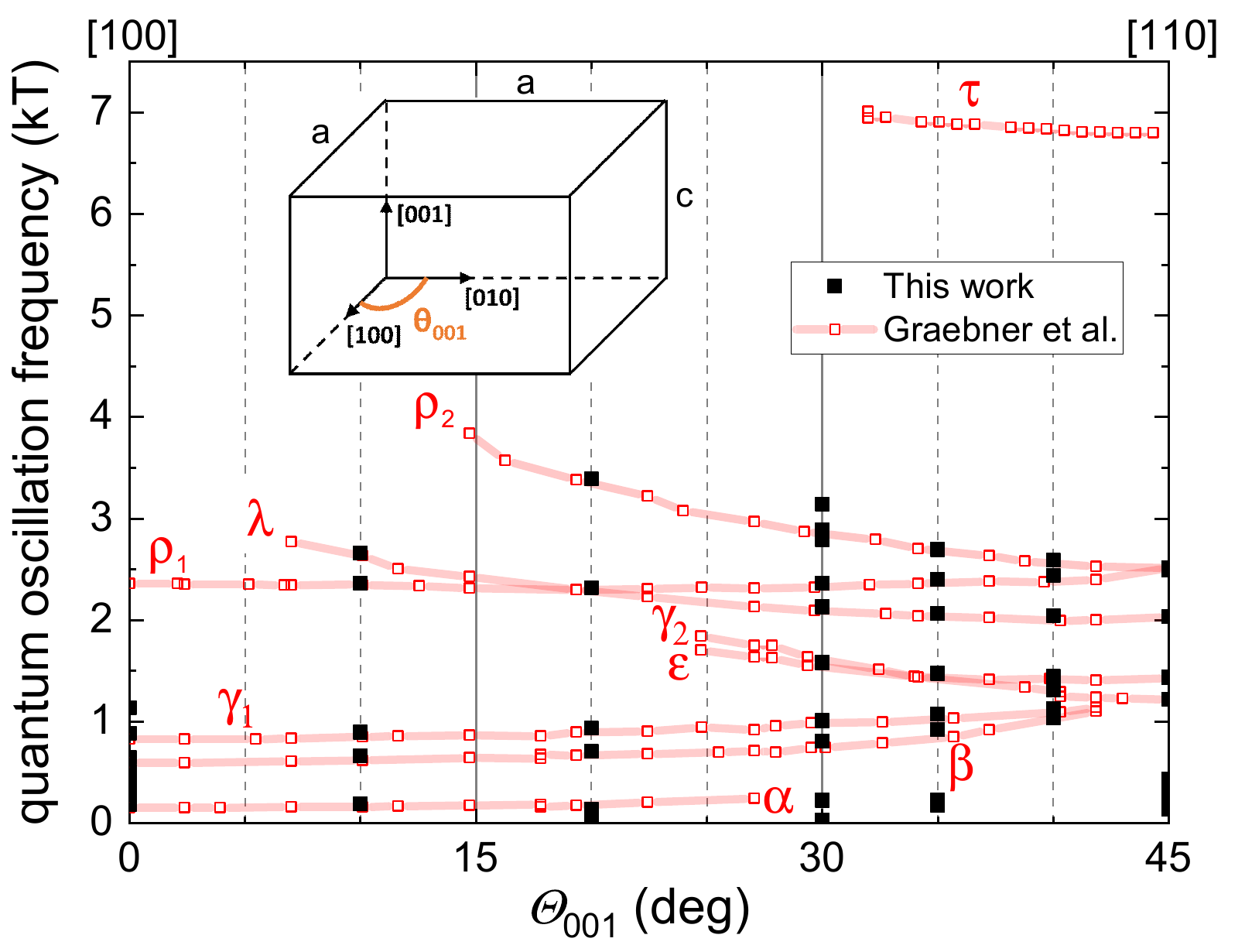}}
	\end{center}
	\vspace{-0mm}
	\caption[Angular dependence of QO frequencies for IrO2.]{Angular dependence of quantum oscillation frequencies of IrO$_2$ for rotation from $B\parallel [100]$ towards $B\parallel [110]$. Full black squares: this work; empty red squares: experimental results from \cite{graebner_magnetothermal_1976}.  Semi-transparent lines are guides to the eye to better illustrate the different branches. The inset defines the rotation angle $\Theta_{001}$.}
	\vspace{-0mm}
	\label{fig:IrO2_angle}
\end{figure}

An example of the measured torque signal at 1.37~K at $\Theta_\mathrm{001}=30^\circ$ is shown in Fig.~\ref{fig:IrO2_example}\,(a). $\Theta_{001}$ is the angle for rotation from $B\parallel [100]$ to $B\parallel [110]$ defined as per the inset to Fig.~\ref{fig:IrO2_angle}. The frequency spectrum of the oscillating part of the signal obtained by fast Fourier transformation (FFT) can be seen in Fig.~\ref{fig:IrO2_example}\,(b) for different temperatures.
We observe a large number of peaks, some of which are harmonics or combination frequencies. The seven most prominent peaks between 225 and 2980~T were identified to be fundamental frequencies coinciding with results in \cite{graebner_magnetothermal_1976}, and were, therefore, labeled with the same Greek letters as in \cite{graebner_magnetothermal_1976}. In addition, we observed a low frequency of about 50~T with a very low FFT amplitude, marked with an asterisk $*$ in Fig.~\ref{fig:IrO2_example}\,(b).

The angular dependence of the dHvA frequencies for IrO$_2$ from this work is displayed as black squares in Fig.~\ref{fig:IrO2_angle} together with the quantum oscillation frequencies found  from magnetothermal measurements~\cite{graebner_magnetothermal_1976} as open red squares. Semi-transparent lines serve as guides for the eye to identify the different branches. We have not included the harmonics or combination frequencies in this display, as they likely stem from interaction effects and not from electronic properties of the sample.

We find that there is a very good correspondence between our experimental results and those from \cite{graebner_magnetothermal_1976}. We observed all known frequency branches, with the exception of the high frequency $\tau$ branch, indicating comparable sample quality in both studies. At some angles, we also observed the low frequency $F^*$, with FFT peaks below the $\alpha$-branch at about 30--50~T. This frequency does not vary with angle but due to the small number of points and the limited frequency resolution, it is not possible to draw firm conclusions about the shape of this FS.

\begin{figure*}[t]
	\begin{center}
		\subfigure{\includegraphics[trim={3.8cm 10cm 4cm 6cm},clip,width=0.45\textwidth]{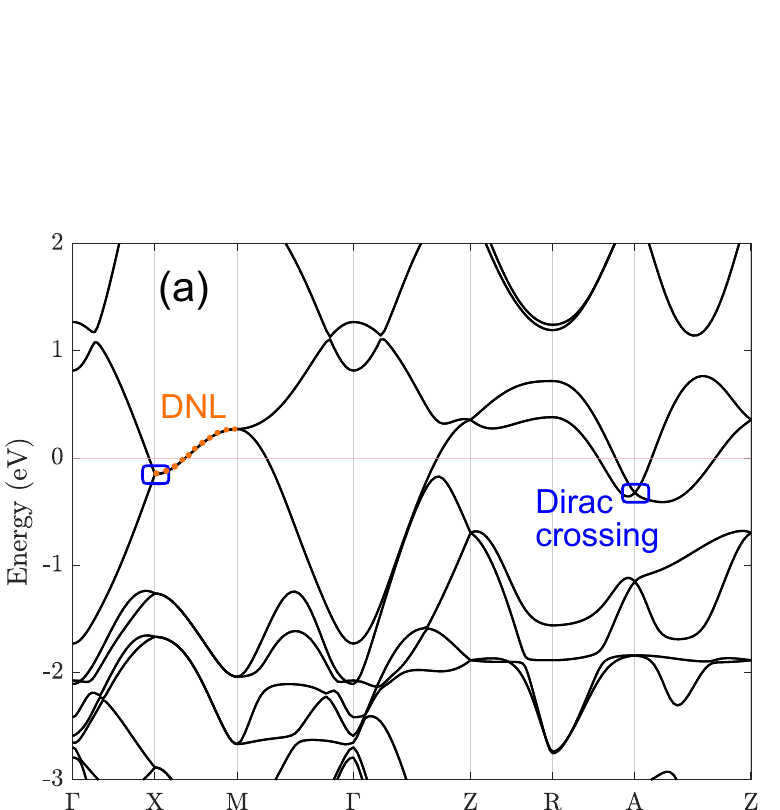}}
			\subfigure{\includegraphics[trim={8.2cm 12.5cm 7cm 0},clip,width=0.54\textwidth]{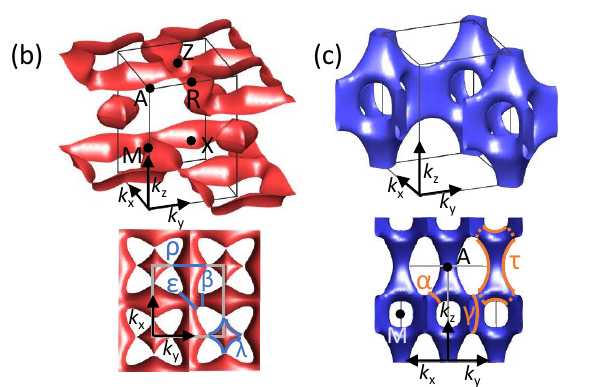}}
	\end{center}
	\vspace{-0mm}
	\caption[Band dispersion with SOC.]{(a) Band dispersion of IrO$_2$ with spin-orbit coupling. One of the DNLs (between \textit{X} and \textit{M}) is highlighted by an orange dashed line and the two Dirac crossings around $X$ and $A$ are marked with blue rectangles. (b) Top: The hole-like Fermi surface and the Brillouin zone including high symmetry points. Bottom: View along $k_z$ for a better visualization of the cross-like structure. (c) Top: The electron-like Fermi surface and the Brillouin zone. Bottom: View along the $[110]$ direction which corresponds to $\Theta_{001}$=45$^\circ$ in our experimental results. For both (b) and (c) bottom figures the extremal orbits for the dHvA frequencies described in the text are drawn and labeled. In both top figures a quarter of the FS in the BZ has been omitted for clarity.}
	\vspace{-0mm}
	\label{fig:DFT}
\end{figure*}

The energy band dispersion of IrO$_2$ as a result of density functional theory (DFT)-based band-structure calculations is displayed in Fig.~\ref{fig:DFT}\,(a) and we find that our results agree well with those in \cite{sun_dirac_2017,xu_strong_2019} and \cite{nelson_dirac_2019} including the DNL between \textit{X} and \textit{M}. There are two bands crossing the Fermi energy leading to two FSs: a hole-like FS consisting of a cross around the \textit{Z} point of the BZ and an ellipsoid around the \textit{M} point (red in Fig.~\ref{fig:DFT}\,(b)), and an electron-like FS which occupies the perimeter of the BZ with holes around \textit{M} and \textit{R} (blue in Fig.~\ref{fig:DFT}\,(c)).
Following the assignments in \cite{graebner_magnetothermal_1976}, we highlight the extremal orbits of the observed dHvA frequencies in the bottom figures of Fig.~\ref{fig:DFT}\,(b) and (c). Frequencies $\beta$, $\epsilon$ and $\rho_{1,2}$ originate from the cross, whereas $\lambda$ is associated with the ellipsoid around \textit{M}. The remaining frequencies $\alpha$, $\gamma_{1,2}$ and $\tau$ stem from the electron FS: $\alpha$ comes from a small neck of this FS (between \textit{M} and \textit{R}). $\gamma_{1,2}$ and $\tau$ can be traced back to the FS portion around \textit{X} and the dog-bone-shaped section whose orbit extends along $k_z$ in the corners of the BZ, respectively, and show a stronger angular dependence \cite{graebner_magnetothermal_1976}.

A fit function based on the Lifshitz-Kosevich (LK) formula \cite{LuttingerWard,LifshitzKosevich,shoenberg} with the temperature damping factor $R_T=X/\sinh(X)$ with $X=14.69 m^*T/\overline{B}$ is applied to the FFT amplitudes of a certain frequency for different temperatures $T$ yielding the effective mass $m^*$ as one of the fit parameters \cite{shoenberg}. Here $\overline{B}^{-1}$ is the center of the inverse field window used in the FFT.
The inset of Fig.~\ref{fig:IrO2_example}\,(b) shows the FFT amplitudes and the curve of the corresponding LK fit function for $F_\lambda=2.16~\mathrm{kT}$ with an effective mass of 2.3(5)~$m_e$. For the other fundamental frequencies we find effective masses between 1.9 and 3.0~$m_e$. $F^*$ has a mass of about 1~$m_e$. The effective masses of the harmonics and combination frequencies in Fig.~\ref{fig:IrO2_example}\,(b) could not be determined reliably due to the limited field and temperature range. The effective masses of all main frequencies observed at $\Theta_\mathrm{001}=30^\circ$ are listed in Table~\ref{tab:IrO2_meff}.

\begin{table}\centering
	\begin{spacing}{1.3} 
	\begin{tabular}{c|c|c|c|c|c}
		$F$ (kT)&  & $m^*(m_e)$ & $T_\mathrm{D}$ (K)& $\tau$ (ps) & $\mu_q$ ($\frac{\mathrm{cm}^2}{\mathrm{V}\cdot\mathrm{s}^2}$)\\[.5ex] \hline
		0.05	&	$F$*	&	1.0(2)	&	4.5(5)	&	0.27(3)	&	500(100)	\\
		0.225	&	$\alpha$	&	2.15(5)	&	0.95(3)	&	1.28(4)	&	1044(41)	\\
		0.77	&	$\beta$	&	2.1(3)	&	0.30(1)	&	4.0(1)	&	3390(500)	\\
		0.98	&	$\gamma_1$	&	1.9(1)	&	0.95(5)	&	1.35(7)	&	1250(100)	\\
		1.64	&	$\gamma_2$	&	2.1(2)	&	1.10(5)	&	1.10(5)	&	920(100)	\\
		2.16	&	$\lambda$	&	2.3(5)	&	1.0(1)	&	1.2(1)	&	930(220)	\\
		2.34	&	$\rho_1$	&	2.0(4)	&	1.3(1)	&	0.93(7)	&	820(175)	\\
		2.98	&	$\rho_2$	&	3.0(3)	&	0.7(1)	&	1.7(2)	&	1020(180)	\\
		
	\end{tabular}
\end{spacing}
	\caption{Experimental values for fundamental frequencies $F$, effective masses $m^*(m_e)$, Dingle temperatures $T_\mathrm{D}$, relaxation times $\tau$ and quantum mobilities $\mu_q$ of dHvA oscillations observed in IrO$_2$ for $\Theta_{001}=30^\circ$.}
	\label{tab:IrO2_meff}
\end{table}

\setstretch{1.0}

\begin{figure}[t]
	\begin{center}
		{\includegraphics[width=.99\columnwidth]{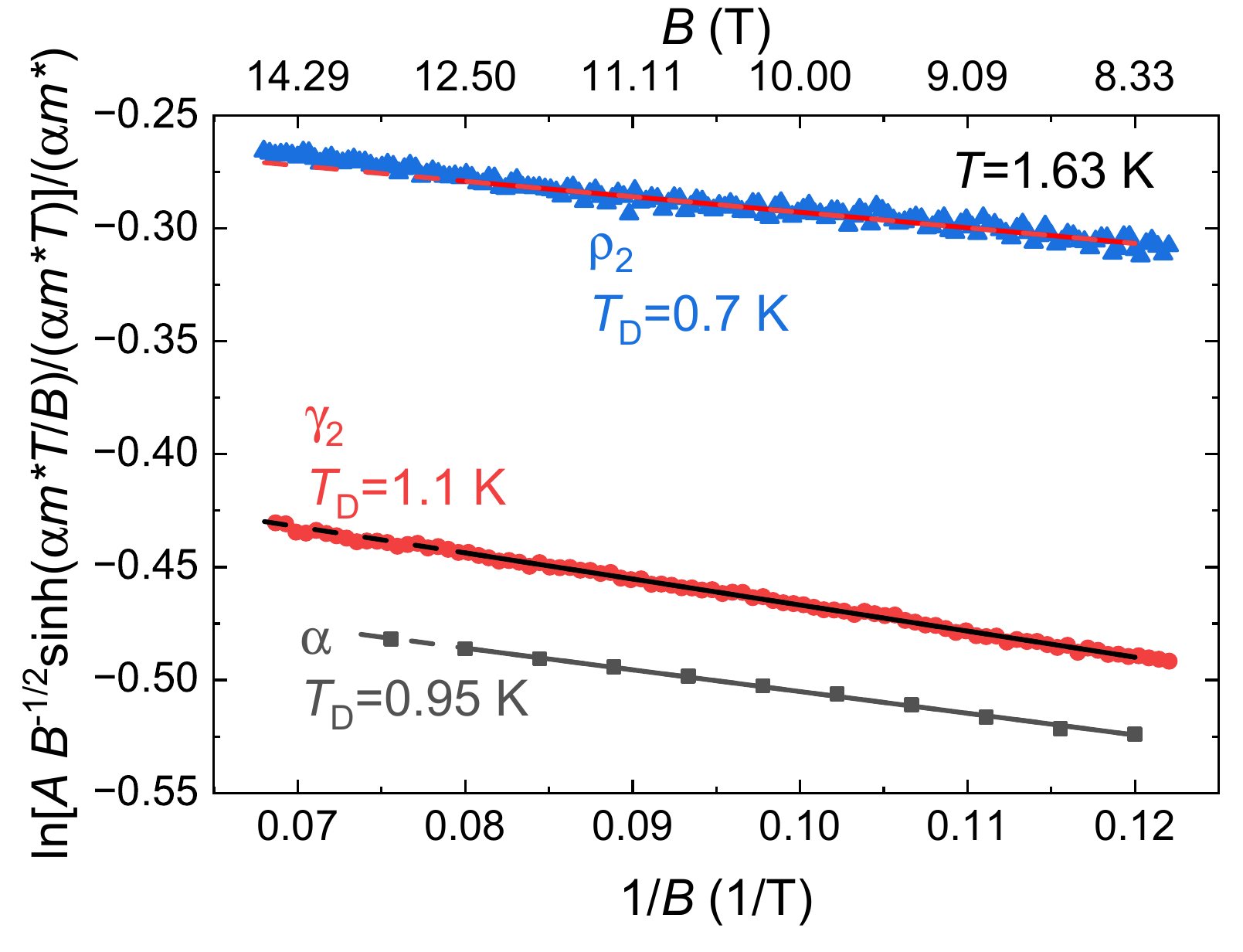}}
	\end{center}
	\vspace{-0mm}
	\caption[Dingle temperature.]{Dingle plot of the field dependent FFT amplitudes at 1.63~K for frequencies $\alpha$, $\gamma_2$ and $\rho_2$ between 6.3 and 15~T. Fit functions were applied between 8.3 and 12.5~T (solid lines). Dashed lines are extrapolations to higher fields (lower 1/$B$) to illustrate that the amplitudes of frequencies $\alpha$ and $\gamma_2$ continue to follow the linear behavior with the same slope whereas the slope for $\rho_2$ shows small deviations.}
	\vspace{-0mm}
	\label{fig:dingle}
\end{figure}

The LK formula can also be applied to the field dependence of the quantum-oscillation amplitudes measured by the torque technique via the so-called Dingle factor $R_\mathrm{D} = \exp(-14.69~\mathrm{T/K} m^* T_\mathrm{D}/B)$ \cite{dingle1952} which depends on the effective mass $m^*$ and the Dingle temperature $T_\mathrm{D}$.
The Dingle temperature is inversely proportional to the relaxation time $\tau$ by $T_\mathrm{D} = \hbar/(2\pi k_\mathrm{B} \tau)$ and is, through its relation to the relaxation time, a measure for sample quality.
The Dingle plot in Fig.~\ref{fig:dingle} displays the amplitude $A$ of the dHvA oscillations of frequencies $\alpha$, $\gamma_2$ and $\rho_2$ at 1.63~K as $\ln(A/[R_T B^{3/2}])$ vs 1/$B$. The Dingle temperature can then be determined from the slope of a linear fit to the data. We have limited the fits to a field range of 8.3 to 12.5~T, similar to the range for the effective mass fits. The dHvA amplitudes in the Dingle plots of $\alpha$, $\gamma_2$ and $\rho_2$ show the expected linear dependence and the low Dingle temperature values of 0.7 to 1.1~K indicate very high sample quality. The Dingle temperatures and relaxation times are included in Table~\ref{tab:IrO2_meff} for all observed frequencies.

We note here that the Dingle plots in Fig.~\ref{fig:dingle} for $\alpha$ and $\gamma_2$ display linear behavior up to the highest measured field of 15~T. The amplitudes of $\rho_2$ show a slight deviation from linear behavior at high fields possibly due to the onset of interaction effects at high oscillation amplitudes. 

We also calculated the quantum mobilities $\mu_q=q \tau/m^*$ from the electron charge~$q$, relaxation time~$\tau$ and the effective mass~$m^*$. 
They are listed in the final column of Table~\ref{tab:IrO2_meff} and range from 500 to 3390~cm$^2$/(Vs$^2$).
 Those are typical values for non-topological electrons in line with the masses and Dingle temperatures observed in this study. Much higher values would only be expected for very light Dirac electrons (e.g. $\mu_q\approx$15000~cm$^2$/(Vs$^2$) for two frequencies in the Dirac node arc semimetal PtSn$_4$ \cite{wang_2018}). We will discuss in the following why such light electrons have not yet been observed in quantum oscillation studies of IrO$_2$ despite the prediction of DNLs and their observation in ARPES experiments.

\section{Discussion}

In our dHvA quantum oscillation experiments we have reproduced almost all of the frequencies and, therefore, the FS cross-sections observed earlier in \cite{graebner_magnetothermal_1976}. In addition, we measured the temperature dependence of the oscillation amplitudes at $\Theta_{001}=30^\circ$ and derived the effective masses, Dingle temperatures and quantum mobilities. Our experimental study is complemented by DFT band-structure calculations that allow us to place the experimentally observed dHvA frequencies in context with the calculated FS topology.

The effective masses are enhanced by a factor of 2--3 compared to the electron rest mass. Those values are comparable to the effective masses in elemental iridium where the highest experimentally observed mass is 2.99~$m_e$ \cite{Hornfeldt1973}. Ir has a Sommerfeld coefficient of $\gamma=3.19~\mathrm{mJ/(mol~K^2)}$ \cite{gubser1973}, whereas IrO$_2$ has a $\gamma$ value of $5.51~\mathrm{mJ/(mol~K^2)}$ \cite{passenheim_heat_1969}. This indicates that even higher masses could be expected in IrO$_2$. A candidate frequency for high masses is the $\tau$ frequency ($>$7~kT) which we have not observed, possibly due to an unfavorable combination of enhanced mass and a curvature factor that causes an additional dampening of those high frequency oscillations.
Also, the ratio between the FFT amplitudes for different frequencies can differ for different techniques \cite{myers_1999} which could explain why the $\tau$ frequency was observed in the magnetothermal oscillations of Ref.~\cite{graebner_magnetothermal_1976} and not in our torque magnetometry data.
While some higher masses could be expected from the specific heat coefficient, very low masses of the order of 0.1~$m_e$ or lower are typically associated with Dirac electrons. DNLs have been predicted for the IrO$_2$ band structure and have also been experimentally observed in ARPES measurements. Such low masses have not been observed in our experiment and in the following we will discuss the most likely reason why this was not the case and consider further experiments that might help to reveal or exclude the existence of Dirac electrons in IrO$_2$.

For comparison, we will review the case of ReO$_2$, a close relative of IrO$_2$ with a similar $\gamma$-value of $3.0~\mathrm{mJ/(mol~K^2)}$ \cite{hirai_extremely_2021}, which was predicted to host Dirac nodal chains \cite{wang_hourglass_2017} and which has been studied using the dHvA and Shubnikov-de Haas (oscillations in electrical transport) effect \cite{Hirai_2023_FSReO2}.

ReO$_2$ was found to have four large FS sections with quantum oscillation frequencies of 100 to 3600~T. Their respective effective masses are of the order of 0.67 to 1.74~$m_e$. So far, those properties are comparable to our findings for IrO$_2$. However, ReO$_2$ also has a very small FS with frequencies of 8--12~T and an extremely low mass of 0.06~$m_e$. This FS is located inside a Dirac loop and, therefore, its low mass is very likely related to the hourglass nodal chains with Dirac points and linear band dispersion.
As mentioned, such low masses have not been observed in our experimental study of IrO$_2$.
The low frequency $F^*$ at about 30--50 T marked in Fig.~\ref{fig:IrO2_example}\,(b) might be close to the low frequency in ReO$_2$ and was observed up to 4.3~K but the result of the LK fit gave an estimated effective mass of the order of 1.0(2)~$m_e$ which is significantly heavier than the very light mass observed in ReO$_2$.

The Dirac points in ReO$_2$ lie rather close to the Fermi energy $E_\mathrm{F}$ (calculated at -105 meV and 136 meV, respectively \cite{Hirai_2023_FSReO2}) and one of the associated nodal chains encircles the aforementioned small FS around the $U$ point of the BZ, making the Dirac electrons observable in quantum oscillation experiments. In IrO$_2$, in turn, the Dirac crossings at $X$ and $A$ sit at -250 and -400~meV, respectively, below $E_\mathrm{F}$ as calculated by DFT (see Fig.~\ref{fig:DFT}\,(a) and \cite{sun_dirac_2017}), and were observed experimentally by ARPES at about -150 and -400~meV by \cite{xu_strong_2019} and at about -200 and below -400~meV by \cite{nelson_dirac_2019}. The DNLs run along the edges of the BZ (see Fig.~2 in \cite{nelson_dirac_2019} for a clear illustration).
Their position further below $E_\mathrm{F}$ makes the observation of very light Dirac electrons by dHvA oscillations less likely in IrO$_2$.
However, the material's very low residual resistivity at low temperatures attributed to the ultrahigh mobility of Dirac electrons \cite{Hirai_2023_FSReO2} hints towards the existence of light electrons that contribute to the material's properties at low temperatures.
As a reference, $\rho$(2K) in ReO$_2$  is 206~n$\Omega$cm \cite{Hirai_2023_FSReO2}. Exact values for IrO$_2$ are not published but can be estimated to range between 145 and 220~n$\Omega$cm from the data in \cite{lin_low_2004} and \cite{ryden_magnetic_1970}. 
This low residual resistivity implies that Dirac electrons and their topological properties play a role for the electronic characteristics of IrO$_2$ and indicates that further experimental effort is warranted.
The low masses of Dirac electrons are typically associated with low frequencies. As the magnetic background of the torque signal requires the subtraction of a higher order polynomial at some angles, there is the risk of unintentionally subtracting low frequency portions of the signal. When searching for low frequencies it might, therefore, be necessary to use different experimental techniques with less pronounced background like magnetoresistance measurements or different magnetometry techniques. The lowest frequency in ReO$_2$, for example, was only detected in Shubnikov-de Haas oscillations of magnetoresistance and not in the torque signal \cite{Hirai_2023_FSReO2}. In addition, such experiments need to be conducted at intermediate temperatures where high frequencies are less dominant.

\section{Summary and Conclusion}

We have performed dHvA experiments on single crystals of the DNL material IrO$_2$ using torque magnetometry with a piezo-resistive microcantilever and have performed DFT-based calculations on the electronic band-structure of this material. 
We find very good correspondence between our experimental data and results in \cite{graebner_magnetothermal_1976} for the angular dependence of the dHvA frequencies, as well as very good agreement for the band dispersion and FS from our calculations and \cite{sun_dirac_2017,xu_strong_2019,nelson_dirac_2019}. The temperature and field dependence of the oscillation amplitudes revealed enhanced effective masses of 2--3 times the electron rest mass and low Dingle temperatures indicating excellent sample quality. 
We did not observe the very low electron masses expected to be associated with DNLs probably because the corresponding band-crossing points sit well below the Fermi energy. However, the low temperature transport properties of IrO$_2$ hint towards the contribution of low mass electrons to the electronic properties of this material and further experiments with complementary techniques are warranted.

\section*{Acknowledgments}

We thank J.~Klotz, T.~Förster, H.~Rosner and J.~Singleton for useful discussions and thank A.~Julian, P.~Ruddy and T.~Orton for technical assistance. Work performed at the University of Warwick is supported by the European Research Council (ERC) under the European Union's Horizon~2020 research and innovation program (Grant Agreement No.~681260). K.G. acknowledges support from DESY Hamburg, Germany, a member of the Helmholtz Association HGF. M.J.P. thanks the Engineering and Physical Sciences Research Council (EPSRC) for additional funding. D.P. acknowledges the EPSRC, UK grant number EP/R024278/1 and the Oxford-ShanghaiTech collaboration project for financial support.
S.N. and J.-R.S. acknowledge the National Supercomputing Centre (NSCC), Singapore, for providing the computational resources. 
For the purpose of open access, the author has applied a Creative Commons Attribution (CC-BY) licence to any Author Accepted Manuscript version arising from this submission.

\section*{Data Availability Statement}

The data that support the findings of this study are openly available: \cite{data}.

\appendix

\section{Magnetic and torque interaction}

Magnetic interaction (MI) is the result of an induced magnetic field in the sample caused by the oscillating magnetization. The effective field inside the sample would then be $B_\mathit{eff}=\mu_0(H_\mathrm{ext}+M_\mathrm{osc})$.
This can lead to a stronger contribution from harmonics and modulate the amplitudes of oscillations.
If more than one frequency is present, a feedback effect causes the occurrence of combination frequencies and their harmonics \cite{shoenberg}.

Torque interaction (TI) occurs due to feedback effects of the oscillating magnetic moment on the position of the lever in the field. This can alter the shape of the oscillations and lead to a stronger contribution of harmonics and side bands \cite{bergemann_quantum_1999}. 
It is highly likely that either MI or TI or a combination of both is responsible for both saw-tooth shaped oscillations and for the observation of prominent combination frequencies in the quantum oscillation spectrum of IrO$_2$ at high fields and very low temperatures. 

\bibliography{IrO2_OsO2_dHvA_lit}

\end{document}